\documentclass{IEEEcsmag}

\usepackage[colorlinks,urlcolor=blue,linkcolor=blue,citecolor=blue]{hyperref}

\usepackage{tabularx}
\usepackage{upmath}
\usepackage{amssymb}
\usepackage{amsmath}
\usepackage{array}
\usepackage{arydshln}
\usepackage{listings, multicol}
\usepackage{filecontents}
\usepackage{parcolumns}
\usepackage{booktabs}

\newcommand\blfootnote[1]{%
  \begingroup
  \renewcommand\thefootnote{}\footnote{#1}%
  \addtocounter{footnote}{-1}%
  \endgroup
}

\newcolumntype{P}[1]{>{\centering\arraybackslash}p{#1}}

\jvol{XX}
\jnum{XX}
\paper{8}
\jmonth{Sep/Oct}
\jname{Computing in Science and Engineering}
\pubyear{2022}

\begin{document}

\sptitle{Department: Scientific Programming}
\editor{Editors: Anshu Dubey, adubey@anl.gov; Konrad Hinsen, konrad.hinsen@cnrs.fr}

\title{PyExaFMM: an exercise in designing high-performance software with Python and Numba}

\author{S. Kailasa}
\affil{Department of Mathematics, University College London}

\author{T. Wang}
\affil{Department of Mechanical and Aerospace Engineering, The George Washington University}

\author{\text{L}. A. Barba}
\affil{Department of Mechanical and Aerospace Engineering, The George Washington University}

\author{T. Betcke}
\affil{Department of Mathematics, University College London}

\markboth{Department Head}{Paper title}

\begin{abstract}
   Numba is a game-changing compiler for high-performance computing with Python. It produces machine code that runs outside of the single-threaded Python interpreter and that fully utilizes the resources of modern CPUs. This means support for parallel multithreading and auto-vectorization if available, as with compiled languages such as C++ or Fortran. In this article we document our experience developing PyExaFMM, a multithreaded Numba implementation of the Fast Multipole Method, an algorithm with a non-linear data structure and a large amount of data organization. We find that designing performant Numba code for complex algorithms can be as challenging as writing in a compiled language.
\end{abstract}

\maketitle
\chapterinitial{Python}\footnote{We refer here to CPython, the popular C-language implementation of Python dominant in computational science.}is designed for memory safety and programmer productivity. Its simplicity allows computational scientists to spend more time on science, and less time tackling software quirks, memory errors, and incompatible dependencies, which all conspire to drain productivity.\blfootnote{For the purpose of open access, the author has applied a Creative Commons Attribution (CC BY) license to any Author Accepted Manuscript version arising} 
The downside is that code runs through an interpreter and is restricted to a single thread via a software construction called the Global Interpreter Lock (GIL). Libraries for high-performance computational science can bypass the GIL by using Python's C interface to call extensions built in C or other compiled languages, which can be multithreaded or compiled to target special hardware features. Popular examples of this approach include NumPy and SciPy, which have together helped to propel Python's popularity in computational science by providing high-performance data structures for numerical data as well as interfaces for compiled implementations of common algorithms for numerical linear algebra, differential equation solvers, and machine learning, among others.
As the actual number crunching happens outside of the interpreter, the GIL only becomes a bottleneck to performance if a program must repeatedly pass control between the interpreter and non-Python code. This is  typical when an optimized compiled language implementation of your desired algorithm doesn't exist in the Python open-source ecosystem, or if a lot of data organization needs to happen within the interpreter to form the input for an optimized NumPy or SciPy code. Previously, an unlucky developer would have had to tackle these issues by writing a compiled-language implementation and connecting it to their Python package, relegating Python's role to an interface. Many computational scientists may lack the software skills or interest in developing and maintaining complex codebases that couple multiple languages.

This scenario is where Numba comes in \cite{Lam2015}. It is a compiler that targets and optimizes Python code written with NumPy's $n$-dimensional array data structure, the ndarray. Its power derives from generating  compiled code optimized for multithreaded architecture from pure Python. Numba promises the ability to develop applications that can rival C++ or Fortran in performance, while retaining the simplicity and productivity of working in Python. We put this promise to the test with PyExaFMM, an implementation of the three-dimensional kernel-independent fast multipole method (FMM) \cite{Ying2004,Greengard1987}. PyExaFMM is open-source,\footnote{\url{https://github.com/exafmm/pyexafmm}} as are the scripts and Jupyter notebooks used to run the experiments in this paper.\footnote{\url{https://github.com/betckegroup/pyexafmm-cise}} Efficient implementations of this algorithm are complicated by its reliance on a tree data structure and a series of operations that each require major data organization and careful memory allocation. These features made PyExaFMM an excellent test case to see whether Numba could free us from the complexities of developing in low-level languages.

We begin with an overview of Numba's design and its major pitfalls. After introducing the data structures and computations involved in the FMM, we provide an overview of how we implemented our software's data structures, algorithms, and application programming interface (API) to optimally use Numba.

\section{BRIEF OVERVIEW OF NUMBA}

Numba is built using the LLVM compiler infrastructure\footnote{\url{https://llvm.org/}} to target a subset of Python code using ndarrays. LLVM provides an API for generating machine code for different hardware architectures such as CPUs and GPUs and is also able to analyze code for hardware-level optimizations such as auto-vectorization, automatically applying them if they are available on the target hardware \cite{Lattner2004}. LLVM-generated code may be multithreaded, bypassing the GIL.  Numba uses the metadata provided by ndarrays describing their dimensionality, type, and layout to generate code that takes advantage of the hierarchical caches available in modern CPUs \cite{Lam2015}. Altogether, this allows code generated by Numba to run significantly faster than ordinary Python code, and often be competitive with code generated from compiled languages such as C++ or Fortran. 

From a programmer's perspective, using Numba (at least naively) doesn't involve a significant code rewrite. Python functions are simply marked for compilation with a special decorator; see listings (\ref{code:loop_fusion}), (\ref{code:nested_function}) and (\ref{code:parallel_multithreading}) for example syntax. This encapsulates the appeal of Numba: the ability to generate high-performance code for different hardware targets from Python, letting Numba take care of optimizations, would allow for significantly faster workflows than is possible with a compiled language.

\begin{figure*}
    \centerline{\includegraphics {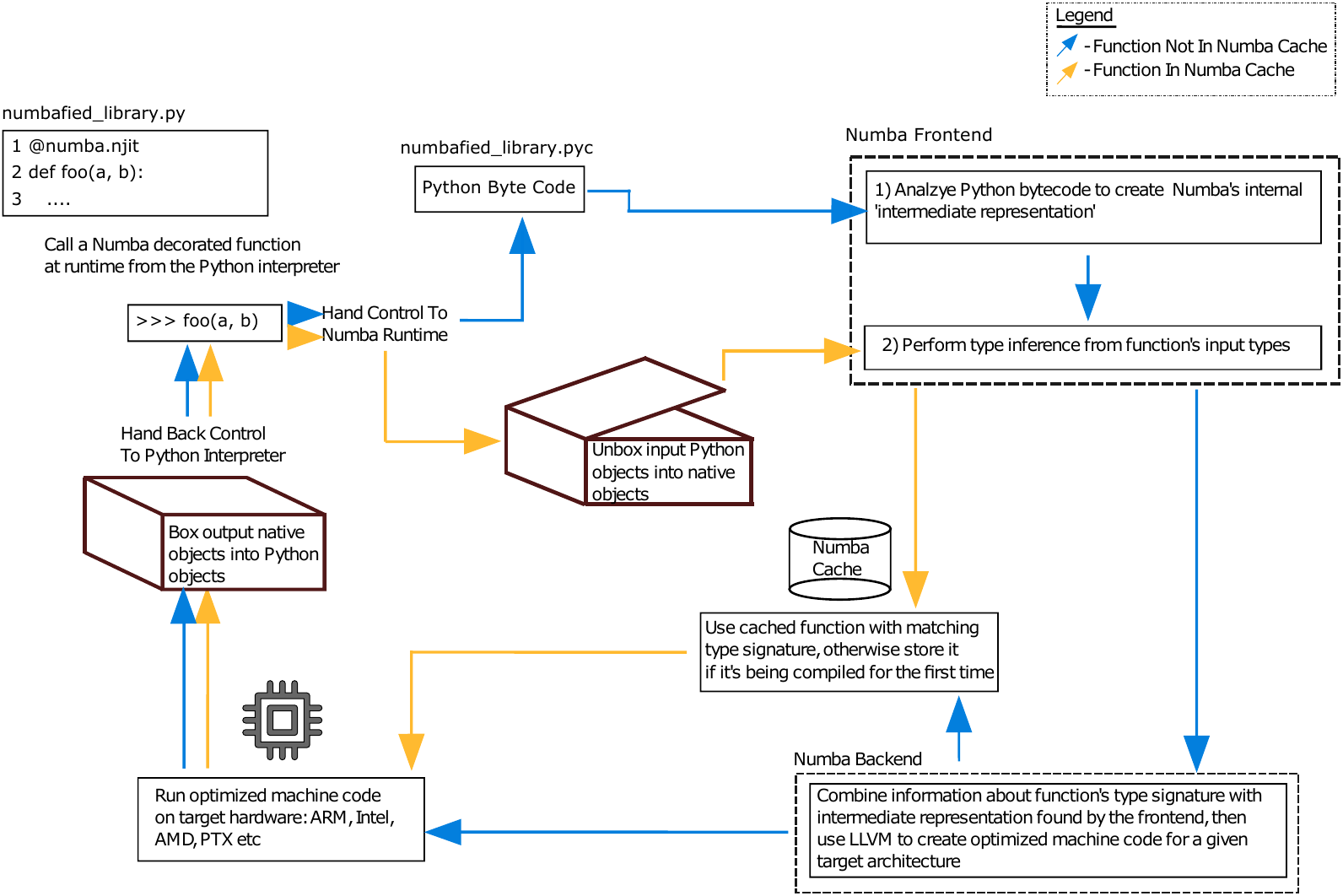}}
    \caption{Simplified execution path when calling a Numba compiled function from the Python interpreter. The blue path is only taken if the function hasn't been called before. The orange path is taken if a compiled version with the correct type signature already exists in the Numba cache.}
    \label{fig:numba}
\end{figure*}

Figure (\ref{fig:numba}) illustrates the program execution path when a Numba-decorated function is called from the Python interpreter. We see that Numba doesn't replace the Python interpreter. If a marked function is called at runtime, program execution is handed to Numba's runtime, which compiles the function on-the-fly with a type signature matching the input arguments. This is the origin of the term `just in time' (JIT) to describe such compilers.

The Numba runtime interacts with the Python interpreter dynamically, and control over program execution is passed back and forth between the two. This interaction is at the cost of having to `unbox' Python objects into types compatible with the compiled machine code, and `box' the outputs of the compiled functions back into Python-compatible objects. This process doesn't involve reallocating memory, however, pointers to memory locations have to be converted and placed in a type compatible with either Numba-compiled code or Python.

\begin{lstlisting}[float=t, caption={An example of using Numba in a Python function operating on ndarrays.}\label{code:loop_fusion}]

import numba
import numpy as np

# Optionally the decorator can take
# the option nopython=True, which
# disallows Numba from running in
# object mode
@numba.jit
def loop_fusion(a):
    # An example of loop fusion, an
    # optimization that Numba is able
    # to perform on a user's behalf
    # when it recognizes that they are
    # operating similarly on a single
    # data structure
    for i in range(10):
        a[i] += 1

    for i in range(10):
        a[i] *= 5

    return a

\end{lstlisting}

\section{PITFALLS OF NUMBA}\label{sec:pitfalls}

Since its first release, Numba has been extended to cover most of NumPy's functionality, as well as the majority of Python's basic features and standard library modules\footnote{A list of supported features for the current release can be found at \url{https://numba.readthedocs.io/en/stable/reference/pysupported.html}}. If Numba is unable to find a suitable Numba type for each Python type in a decorated function, or it sees a Python feature it doesn't yet support, it runs in `object mode', handling all unknown quantities as generic Python objects. To ensure a seamless experience, this is silent to the user, unless explicitly marked to run in `no Python' mode. Object mode is often no faster than vanilla Python, leaving the programmer to understand when and where Numba works. As Numba influences the way Python is written it's perhaps more akin to a programming framework than just a compiler.

An example of Numba's framework-like behavior arises when implementing algorithms that share data, and have multiple logical steps as in listing (\ref{code:nested_function}). This listing shows three implementations of the same logic: the initialization of a dictionary with some data followed by two matrix multiplications, from which a column of each is stored in the dictionary. The runtimes of all three implementations are shown in table (\ref{tab:boxing_inlining}) for different problem sizes \footnote{All experiments in this work were taken on an AMD Ryzen Threadripper 3970X 32-Core processor running Python 3.8.5 and Numba 0.53.0}. This example is designed to illustrate how arbitrary changes to writing style can impact the behavior of Numba code. The behavior is likely due to the initialization of a dictionary from within a calling Numba function, rather than an external dictionary. However, the optimizations taken by Numba are presented opaquely to a user.

\begin{table}
    \caption{ Testing the effect of the different implementations of computing dense matrix-vector products in double precision with some data storage from listing (\ref{code:nested_function}). }
    \label{table}
    \small
    \begin{tabular*}{17.5pc}{@{}p{40pt}p{50pt}<{\raggedright}p{50pt}<{\raggedright}@{}}
    \\
    Algorithm & Matrix Dimension & Time ($\mu$s) \\
    \hline
     1 & $\mathbb{R}^{1 \times 1}$ &       $1.55  \pm 0.01$     \\
      & $\mathbb{R}^{100 \times 100}$ &   $304  \pm 3$  \\
      & $\mathbb{R}^{1000 \times 1000}$ & $29100 \pm 234$  \\
    \hline
     2 & $\mathbb{R}^{1 \times 1}$ &       $2.73  \pm 0.01$     \\
      & $\mathbb{R}^{100 \times 100}$ &   $312   \pm 3$ \\
      & $\mathbb{R}^{1000 \times 1000}$ & $25700 \pm 92$\\
    \hline
     3 & $\mathbb{R}^{1 \times 1}$ &        $2.72  \pm 0.01$ \\
      & $\mathbb{R}^{100 \times 100}$ &    $312  \pm 1$ \\
      & $\mathbb{R}^{1000 \times 1000}$ &  $25700  \pm 140$  \\
      \\
    
    \end{tabular*}
    \label{tab:boxing_inlining}
\end{table}

Furthermore, not every supported feature from Python behaves in a way an ordinary Python programmer would expect, which has an impact on program design. An example of this arises when using Python dictionaries, which are central to Python, but are only partially supported by Numba. As they are untyped and can have any Python objects as members, they don't neatly fit into a Numba-compatible type. Programmers can declare a Numba-compatible `typed dictionary', where the keys and values are constrained to Numba-compatible types, and can pass it to a Numba decorated function at low cost. However, using a Numba dictionary from the Python interpreter is \textit{always slower} than an ordinary Python dictionary due to the (un)boxing cost when getting and setting any item.

Therefore, though Numba is advertised as an easy way of injecting performance into your program via simple decorators, it has its own learning curve. Achieving performance requires a programmer to be familiar with the internals of its implementation and potential discrepancies that arise when translating between Python and the LLVM-generated code, which may lead to significant alterations in the design of algorithms and data structures. 

\begin{lstlisting}[caption={Three ways of writing an algorithm that performs some computations and saves the results to a dictionary.}\label{code:nested_function}]
import numpy as np
import numba
import numba.core
import numba.typed

# Initialize in Python interpreter
data = numba.typed.Dict.empty(
    key_type=numba.core.types.unicode_type,
    value_type=numba.core.types.float64[:]
)


data['initial'] = np.ones(N)

@numba.njit
def step_1(data):
    a = np.random.rand(N, N)
    data['a'] = (a @ a)[0,:]

@numba.njit
def step_2(data):
    b = np.random.rand(N, N)
    data['b'] = (b @ b)[0,:]

@numba.njit
def algorithm1(data):
    step_1(data)
    step_2(data)

@numba.njit
def algorithm2():
    data = dict()
    data['initial'] = np.ones(N)
    step_1(data)
    step_2(data)
    return data

@numba.njit
def algorithm3():
    data = dict()
    data['initial'] = np.ones(N)

    def step_1(data):
        a = np.random.rand(N, N)
        data['a'] = (a @ a)[0,:]

    def step_2(data):
        b = np.random.rand(N, N)
        data['b'] = (b @ b)[0,:]

    step_1(data)
    step_2(data)
    return data
\end{lstlisting}

\section{THE FAST MULTIPOLE METHOD}

The particle FMM is an approximation algorithm for $N$-body problems, in which $N$ source particles interact with $N$ target particles \cite{Greengard1987}. Consider the calculation of electrostatic potentials in 3D, which we use as our reference problem. Given a set of $N$ charged particles with charge $q_i$ at positions $x_i$, the potential, $\phi_j$, at a given target particle at $x_j$ due to all other particles, excluding self-interaction, can be written as

\begin{equation}
    \phi_j = \sum_{i=1, i \neq j}^{N} \frac{q_i}{4 \pi| x_i-x_j |},
    \label{eq:laplace_kernel}
\end{equation}

\noindent where $\frac{1}{4 \pi| x_i-x_j|}$ is called the kernel, or the Green's function. The naive computation over all particles scales as $O(N^2)$; the FMM compresses groups of interactions far away from a given particle using \textit{expansions}, and reduces the overall complexity to $O(N)$. Expansions approximate charges contained within subregions of an octree, and can be truncated to a desired accuracy, defined by a parameter, $p$, called the expansion order. Problems with this structure appear with such frequency in science and engineering that the FMM has been described as one of the ten most important algorithms of the twentieth century \cite{Dongarra2000}.

The FMM algorithm relies on an octree data structure to discretize the problem domain in 3D. Octrees place the region of interest in a cube, or `root node', and subdivide it into 8 equal parts. These `child nodes' in turn are recursively subdivided until a user-defined threshold is reached (see fig. 1 of Sundar et al.\cite{Sundar2007}). The FMM consists of eight operators: P2M, P2L, M2M, M2L, L2L, L2P, M2P and the `near field', applied \textit{once} to each applicable node over the course of two consecutive traversals of the octree (bottom-up and then top-down). The operators define interactions between a given `target' node, and potentially multiple `source' nodes from the octree. They are read as `X to Y', where `P' stands for particle(s), `M' for \textit{multipole expansion} and `L' for \textit{local expansion}. The direct calculation of (\ref{eq:laplace_kernel}) is referred to as the P2P operator, and is used as a subroutine during the calculation of the other operators. The kernel-independent FMM (KIFMM) \cite{Ying2004} implemented by PyExaFMM is a re-formulation of the FMM with a structure that favors parallelization. Indeed, all of the operators can be decomposed into matrix-vector products, or multithreaded implementations of (\ref{eq:laplace_kernel}), which are easy to optimize for modern hardware architectures, and fit well with Numba's programming framework. We defer to the FMM literature for a more detailed discussion on the mathematical significance of these operators \cite{Ying2004,Greengard1987}.

\section{COMPUTATIONAL STRUCTURE OF FMM OPERATORS}

The computational complexities of KIFMM operators are defined by a user-specified $n_{crit}$, which is the maximum allowed number of particles in a leaf node; $n_e$ and $n_c$, which are the numbers of quadrature points on the \textit{check} and \textit{equivalent} surfaces respectively (see sec. 3 of Ying et al.\cite{Ying2004}). The parameters $n_e$ and $n_c$ are quadratically related to the expansion order, i.e., $n_e = 6(p-1)^2 + 2$ \cite{Ying2004}. Typical values for $n_{crit}$ used are $\sim 100$. Notice the depth of the octree is defined by $n_{crit}$, and hence by the particle distribution.

The near field, P2M, P2L, M2P, and L2P operate independently over the leaf nodes. The M2L and L2L operate independently on all nodes at a given level, from level 2 to the leaf level, during the top-down traversal. The M2M is applied to each node during the bottom-up traversal.
All operators, except the M2L M2M and L2L, rely on P2P. The inputs for P2P are vectors for the source and target positions, and the source charges or expansion coefficients; the output is a vector of potentials.
The inputs to the M2L, M2P, P2L and near-field operators are defined by `interaction lists' called the V, W, X and U lists respectively. These interaction lists define the nodes a target node interacts with when an operator is applied to it. We can restrict the size of these interaction lists by demanding that neighboring nodes at the leaf level are at most twice as large as each other \cite{Sundar2007}. Using this `balance condition', the V, X, W and U lists in 3D contain at most 189, 19, 148 and 60 nodes, respectively.

The near-field operator applies the P2P between the charges contained in the target and the source particles of nodes in its U list, in $O(60 \cdot n_{crit}^2)$. The M2P applies the P2P between multipole expansion coefficients of source nodes in the target's W list and the charges it contains internally in $O(148 \cdot n_e \cdot n_{crit})$. Similarly, the L2P applies the P2P between a target's own local expansion coefficients and the charges it contains in $O(n_e \cdot n_{crit})$.

The P2L, P2M and M2L involve creating local and multipole expansions, and rely on a matrix-vector product related to the number of source nodes being compressed, which for the P2L and M2L operators are defined by the size of the target node's interaction lists. These matrix -ector products have complexities of the form $O(k \cdot n_e^2)$ where $k = |X| = 19$ for the P2L, $k = |V| = 189$ for the M2L, and $k = 1$ for the P2M. Additionally, the P2L and P2M have to calculate `check potentials' \cite{Ying2004} that require $O(19 \cdot n_{crit} \cdot n_c)$ and $O(n_{crit} \cdot n_c)$ calculations, respectively. The M2M and L2L operators both involve translating expansions between nodes and their eight children, and rely on a matrix-vector product of $O(n_e^2)$.

The structure of the FMM's operators exposes natural parallelism. The P2P is embarrassingly parallel over each target, as are the M2L, M2P, P2L and near-field operators over their interaction lists. The near-field, L2P, M2P, P2L and P2M operators are also embarrassingly parallel over the leaf nodes, as are the M2L, M2M and L2L over the nodes at a given level.

\section{DATA-ORIENTED DESIGN OF PYEXAFMM}

In the context of high-performance computing, data-oriented design refers to a coding approach that favors data structures with simple memory layouts, such as arrays. The aim is to effectively utilize modern hardware features by making it easier for programmers to optimize for cache locality and parallelization.
In contrast, object-oriented design involves organizing code around user-created types or objects, where the memory layout is complex and can contain multiple attributes of different types. 
This complexity makes it difficult to optimize code for cache locality, and thus it results in lower performance in terms of utilizing hardware features.

Numba focuses on using ndarrays, in alignment with the data-oriented design principles, which we apply in the design of PyExaFMM's octrees as well as its API.
Octrees can be either `pointer based' \cite{Wang2021}, or `linear' \cite{Sundar2007}. A pointer-based octree uses objects to represent each node, with attributes for a unique id, contained particles, associated expansion coefficients, potentials, and pointers to their parent and sibling nodes. This makes searching for neighbors and siblings a simple task of following pointers. The linear octree implemented by PyExaFMM represents nodes by a unique id stored in a 1D vector, with all other data such as expansion coefficients, particle data, and calculated potentials, also stored in 1D vectors. Data is looked up by creating indices to tie a node's unique id to the associated data. This is an example of how using Numba can affect design decisions, and make software more complex, despite the data structures being simpler.

\begin{figure*}
    \centerline{\includegraphics {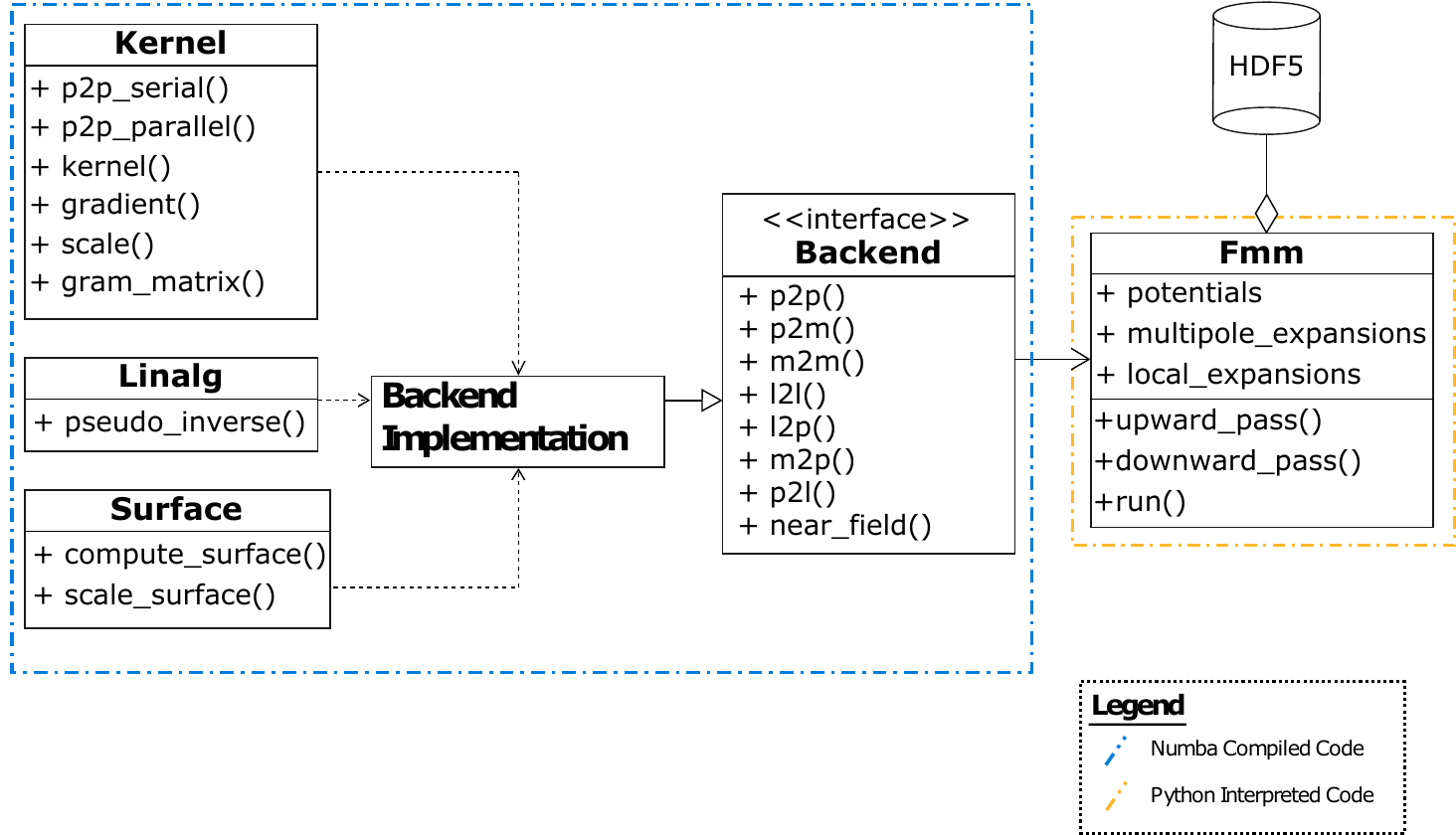}}
    \caption{Simplified UML model of all PyExaFMM components. Trees and other precomputed quantities are stored in an HDF5 database. The \texttt{Fmm} object acts as the user interface; all other components are modules consisting of methods operating on arrays.}
    \label{fig:design}
\end{figure*}

Figure (\ref{fig:design}) illustrates PyExaFMM's design. It has a single Python object, \texttt{Fmm}, acting as the API. It initializes ndarrays for expansion coefficients and calculated potentials, and its methods interface with Numba-compiled functions for the FMM operators and their associated data-manipulation functions. We prefer nested functions when sharing data, but keep the operator implementations separate from each other, which allows us to unit test them individually. This means that we must have at least one interaction between Numba and the Python interpreter to call the near field, P2M, L2P, M2P and P2L operators, $d-2$ interactions to call the  M2L and L2L operators, and $d$ interactions for the M2M operator, where $d$ is the depth of the octree. The most performant implementation would be a single Numba routine that interacts with Python just once, however this would sacrifice other principles of clean software engineering such as modularity, and unit testing.

This structure has strong parallels with software designs that arise from traditional methods of achieving performance with Python by interfacing with a compiled language such as C or Fortran. The benefit of Numba is that we can continue to write in Python. Yet as seen above, performant Numba code may only be superficially Pythonic through its shared syntax.

\section{MULTITHREADING IN NUMBA}

Numba enables multithreading via a simple parallel for-loop syntax (see listing (\ref{code:parallel_multithreading})) reminiscent of OpenMP. Internally, Numba can use either OpenMP or Intel TBB to generate multithreaded code. We choose OpenMP for PyExaFMM, as it's more suited to functions in which each thread has an approximately similar workload. The threading library can be set via the \lstinline{NUMBA_THREADING_LAYER} environment variable.

Numerical libraries like NumPy and SciPy use multithreaded compiled libraries such as OpenBLAS or IntelMKL to execute mathematical operations internally. When these operations are compiled with Numba, they retain their internal multithreading. If this is combined with a multithreaded region declared with Numba, as in listing (\ref{code:parallel_multithreading}), it can lead to \emph{nested parallelism}, where a parallel region calls a function that contains another parallel region inside it.
This creates \emph{oversubscription}, where the number of active threads exceeds the CPU's capacity, resulting in idle threads and broken cache locality, and possibly hanging threads,waiting for others to finish. To avoid this, PyExaFMM explicitly sets NumPy operations to be single-threaded by using the environment variable \lstinline{OMP_NUM_THREADS=1} before starting the program. This ensures that only the threads declared using Numba are created.

\begin{lstlisting}[float=t, caption={An example of parallel multithreading.}\label{code:parallel_multithreading}]
import numba
import numpy as np

@numba.njit(cache=True, parallel=True)
def multithreading(A):
    # Numpy is configured to run single
    # threaded to avoid thread
    # oversubscription from the interaction
    # between Numba and Numpy threads.
    for _ in numba.prange(10):
            b = a @ a
\end{lstlisting}

\section{PARALLELIZATION STRATEGIES FOR FMM OPERATORS}

The P2M, P2L, M2P, L2P and near-field operators all rely on the P2P operator, as this computes Equation \eqref{eq:laplace_kernel} over their respective sources and targets, parallelized over their targets, the leaf nodes.

For the L2P operator we encourage cache locality for the P2P step, and keep the data structures passed to Numba as simple as possible, by allocating 1D vectors for the source positions, target positions and the source expansion coefficients, such that all the data required to apply an operator to a single target node is adjacent in memory. By storing a vector of index pointers that bookend the data corresponding to each target in these 1D vectors, we can form parallel for-loops over each target to compute the P2P encouraging cache locality in the CPU. In order to do this, we have to first iterate through the target nodes and look up the associated data to fill the cache local vectors.
The speedup achieved with this strategy, in comparison to a naive parallel iteration over the L2P's targets, increases with the number of calculations in each thread and hence the expansion order $p$. In an experiment with 32,768 leaves, $n_{crit} = 150$ and $p=10$, our strategy is $13\%$ faster. This is a realistic FMM setup with approximately $10^6$ randomly distributed particles on the surface of a sphere.

The previous strategy is too expensive in terms of memory for the near-field, M2L and M2P operators, due to their large interaction lists. For example, allocating an array large enough to store the maximum possible number of source particle coordinates in double precision for the M2P operator, with $|W|=148$ and $n_{crit}=150$, requires $\sim 17$GB for the above experiment, and a runtime cost for memory allocations that exceeds the computation time. Instead, for the M2L we perform a parallel loop over the target nodes at each given level, and over the leaf nodes for the M2P and near field, looking up the relevant data from the linear tree as needed. The P2L interaction list of each target is at most 19 nodes, while the P2M must also calculate a check potential, canceling out any speedup from cache locality for these operators.

The matrices involved in the M2M and L2L operators can be precomputed and scaled at each level \cite{Wang2021}, and their application is parallelized over all nodes at a given level.

Multithreading in this way means that we call the P2M, M2P, L2P and near-field operators once during the algorithm, the M2L and L2L are called $d-2$ times, and the M2M is called $d$ times, where $d$ is the depth of the octree. This is the minimum number of calls while keeping the operator implementations separate for unit testing.

\begin{figure}
	\centerline{\includegraphics[width=8cm]{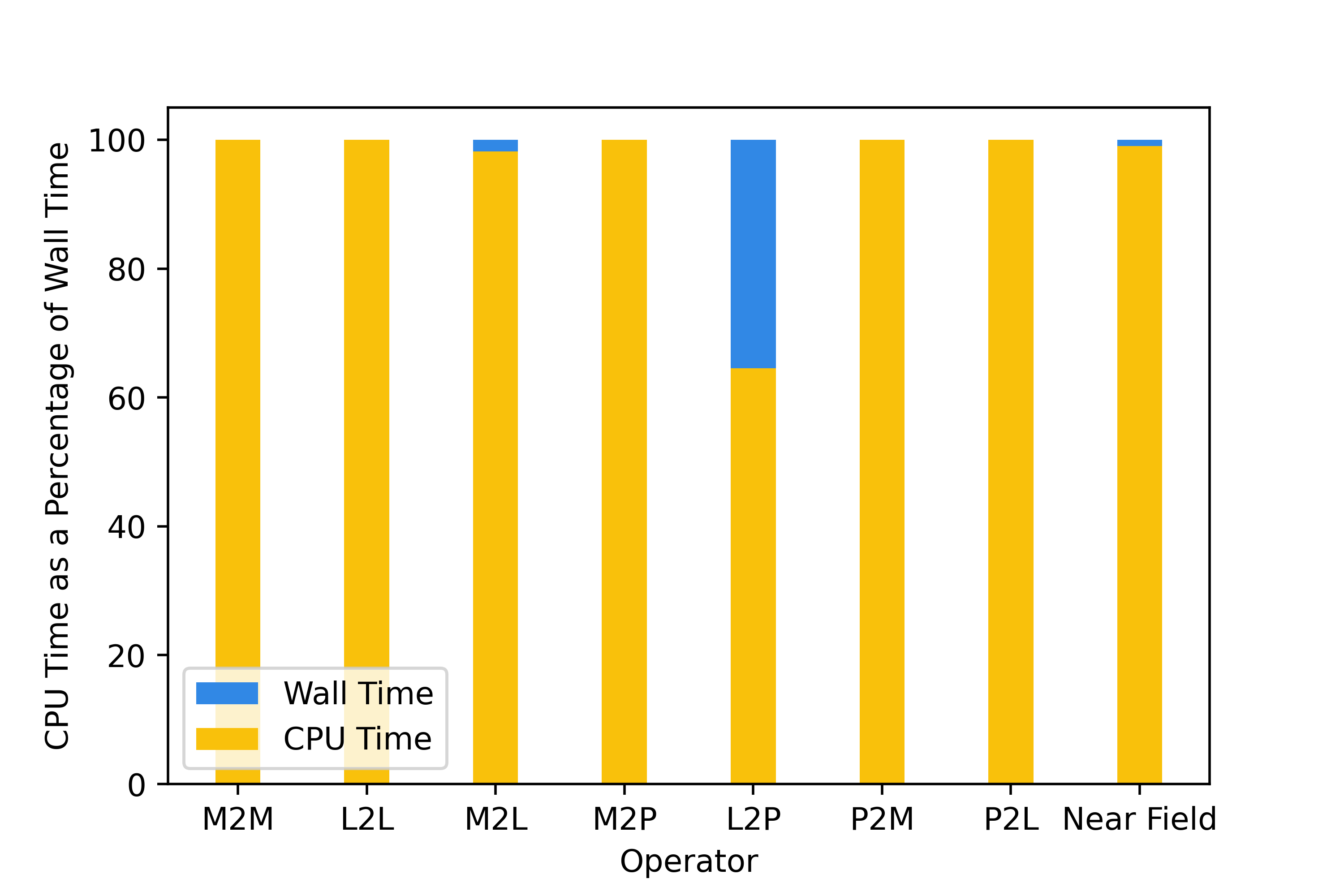}}
    \caption{CPU time as a percentage of wall time for operators. CPU time is defined as the time in which the algorithm runs pure Numba compiled functions. Wall time is CPU time in addition to the time taken to return control to the Python interpreter. } 
	\label{fig:cpu_wall}
\end{figure}

Figure (\ref{fig:cpu_wall}) compares the time spent within each Numba-compiled operator (`CPU time') to the total runtime (`wall time') of each operator. The results are computed over five trials with 32,768 leaves,  $n_{crit}=150$ and $p=6$, for a random distribution of $10^6$ charges distributed on the surface of a sphere, representing a typical FMM setup. The mean sizes of the interaction lists are $|U|=11$, $|V|=42$, $|X|=3$, $|W|=3$, and the entire algorithm is computed in $5.95 \pm 0.02 s$, with an additional $9.00 \pm 0.01 s$ for operator pre-computations for a given dataset---a speed that is unachievable in ordinary single-threaded interpreted Python.

The wall time includes the time to (un)box data, organize inputs for Numba compiled functions, and pass control between Numba and Python. Except for the L2P, which has a different parallelization strategy  requiring significant data organization that must take place within the GIL-restricted Python interpreter, the runtime costs are less than 5\% of each operator's total wall time, implying that we are nearly always running multithreaded code and utilizing all available CPU cores. 

\section{CONCLUSION}

To achieve optimal multithreaded performance with Numba, careful consideration of the algorithm, Numba's backend implementation, and a data-oriented design are required. The pitfalls illustrated above demonstrate the potential need for care when adapting code to achieve performance with Numba. While Numba is marketed as a simple way to enhance performance of Python code with a decorator, implementing complex algorithms requires significant software development expertise. This level of expertise may exceed the capabilities of Numba's target audience.

Nevertheless, Numba is truly a remarkable tool. For projects that prioritize Python's expressiveness, simple cross-platform builds, and a vast open-source ecosystem, having only a few isolated performance bottlenecks, Numba is a game-changer. By writing solely in Python, our PyExaFMM project remains concise, with just 4901 lines of code. Best of all, we can effortlessly deploy it cross-platform with Conda and distribute our software through popular Python channels, avoiding the need to create and maintain a separate Python interface, a tedious and time-consuming task commonly associated with compiled language packages for computational science.
We encourage readers to explore this powerful tool and engage with the Numba community, which continues to push the boundaries of high-performance computing in the Python ecosystem.

\section{ACKNOWLEDGMENT}

SK is supported by EPSRC Studentship 2417009. TB is supported by EPSRC Grants EP/W026260/1 and EP/W007460/1.

\bibliography{pyexafmm}

\bibliographystyle{ieeetr}

\begin{IEEEbiography}{Srinath Kailasa}{\,}is a PhD student in Mathematics at University College London. Contact him at srinath.kailasa.18@ucl.ac.uk.
\end{IEEEbiography}

\begin{IEEEbiography}{Tingyu Wang}{\,} has a PhD in Mechanical Engineering from the George Washington University, and is a Senior Software Engineer at NVIDIA. Contact him at tingyu.wang6@gmail.com.
\end{IEEEbiography}

\begin{IEEEbiography}{Lorena. A. Barba}{\,} is a Professor of Mechanical and Aerospace Engineering in the School of Engineering and Applied Science at the George Washington University (GWU), Washington, DC. Her research interests include computational fluid dynamics, high-performance computing, aerodynamics and biophysics. Barba received a Ph.D. in Aeronautics from the California Institute of Technology and a B.Sc. and PEng in Mechanical Engineering from Universidad T\'ecnica Federico Santa Mar\`ia in Chile. She is a member of IEEE CS, SIAM, AIAA and ACM. Contact her at labarba@gwu.edu.
\end{IEEEbiography}

\begin{IEEEbiography}{Timo Betcke}{\,} is a Professor of Computational Mathematics in the Department of Mathematics, University College London (UCL), U.K. His research interests include numerical analysis, scientific computing, boundary element methods and inverse problems. Betcke received a DPhil in Numerical Analysis from Oxford University. Contact him at t.betcke@ucl.ac.uk.\end{IEEEbiography}

\end{document}